\pdfoutput=1

\documentclass[epj]{svjour}

\usepackage{graphicx}
\usepackage{epstopdf}

\usepackage{bold-extra}		

\usepackage{subcaption}		
\usepackage[pagebackref=false]{hyperref}
\hypersetup{
  colorlinks = true,
  linkcolor=blue,   
  citecolor=blue,   
  urlcolor=blue,    
  pdfdisplaydoctitle=true
}

\usepackage[numbers, square, comma, sort&compress, merge]{natbib}	

\newcommand{\gosia}{{\rmfamily \textsc{gosia}}}	

\begin{document}
\titlerunning{Analysis methods using the \gosia{} code}
\title{Analysis methods of safe Coulomb-excitation experiments with radioactive ion beams using the \gosia{} code}

\author{
M.~Zieli\'{n}ska\inst{1,}\thanks{\emph{Corresponding author:} magda.zielinska@cea.fr}  \and
L.~P.~Gaffney\inst{2,3} \and 
K.~Wrzosek-Lipska\inst{2,4} \and 
E.~Cl\'{e}ment\inst{5} \and
T.~Grahn\inst{6,7} \and 
N.~Kesteloot\inst{2,8} \and 
P.~Napiorkowski\inst{4} \and 
J.~Pakarinen\inst{6,7} \and 
P.~Van~Duppen\inst{2} \and 
N.~Warr\inst{9}
}                     
%
%
\institute{
CEA Saclay, IRFU/SPhN, F-91191 Gif-sur-Yvette, France \and
KU Leuven, Instituut voor Kern- en Stralingsfysica, B-3001 Leuven, Belgium \and
School of Engineering, University of the West of Scotland, Paisley PA1 2BE, United Kingdom \and
Heavy Ion Laboratory, University of Warsaw, PL-00-681 Warsaw, Poland \and
GANIL, BP-5027, F-14076 Caen Cedex, France \and
University of Jyvaskyla, Department of Physics, P.O. Box 35, FI-40014, University of Jyvaskyla, Finland \and
Helsinki Institute of Physics, P.O.Box 64, FI-00014 University of Helsinki, Finland \and
Belgian Nuclear Research Centre, SCK$\bullet$CEN, 2400 Mol, Belgium \and
Institut f\"{u}r Kernphysik, Technische Universit\"{a}t Darmstadt, D-64289 Darmstadt, Germany
}
\date{Received: date / Revised version: date}
%
\abstract{With the recent advances in radioactive ion beam technology, Coulomb excitation at safe energies
becomes an important experimental tool in nuclear-structure physics. 
The usefulness of the technique to extract key information on the electromagnetic properties of nuclei has been demonstrated since the 1960's with stable beam and target combinations. 
New challenges present themselves when studying exotic nuclei with this technique, including dealing with low statistics or number of data points, absolute and relative normalisation of the measured cross sections and a lack of complementary experimental data, such as excited-state lifetimes and branching ratios.
This paper addresses some of these common issues and presents analysis techniques to extract transition strengths and quadrupole moments utilising the least-squares fit code, \gosia{}.
} 
\maketitle
\section{Introduction} \label{sec:intro}

Recent advances in radioactive ion beam (RIB) technology, in particular the
increasing range of species and post-acceleration energies available from
ISOL facilities such as REX-ISOLDE at CERN, SPIRAL at GANIL and ISAC at
TRIUMF, has led to a resurgence of the use of nuclear reactions to study the
structure of nuclei~\cite{Jenkins2014}.  Specifically, Coulomb excitation at
safe energies with RIBs is now giving us a wide range of information on the
electromagnetic properties of exotic nuclei, leading to knowledge of the
nuclear shape or, more precisely, nuclear charge
distribution~\cite{Gorgen2010}.

``Safe'' Coulomb excitation is defined as the process of inelastic
scattering of nuclei via the electromagnetic force such that the energy in
the centre-of-mass frame ensures negligible contribution to the reaction
process from the strong force.  This is fulfilled by maintaining a minimum
distance of 5~fm between the nuclear surfaces, often called Cline's
``safe energy'' criterion~\cite{Cline1986}.  Exploiting the well-understood
electromagnetic interaction allows a nuclear-model-independent interpretation of the
observed data.  With the use of light-ions as probes, the excitation modes
are often limited to single transitions from the ground state.  This data
can be interpreted in terms of a semi-classical description using
first-order perturbation theory.  However, the use of high-$Z$ probes has
meant that multiple-step excitation is now common, and a large number of
states can be accessed from ground or isomeric states.  The technique of
data analysis based on coupled-channel calculations with the \gosia{}
code~\cite{Czosnyka1983,*GosiaManual} has allowed for the determination of
large, and in some cases complete, sets of low-lying $E2$ and $E3$ matrix
elements in multi-step Coulomb-excitation experiments, including diagonal
matrix elements related to the static electromagnetic moments.  Due to this
completeness of measurement, low-energy Coulomb excitation with heavy ions
(or high-$Z$ targets) is an extremely sensitive probe of collective nuclear
structure.  Used in conjunction with complementary spectroscopic data, such
as excited-state lifetimes, $\gamma$-ray and conversion electron branching
ratios, multipole mixing ratios, electric and magnetic moments, mean-square
charge radii etc., a pure experimental understanding of low-lying collective
modes and shapes can be achieved.

When studying exotic nuclei with this technique however, new challenges emerge. These include dealing with low statistics and a lack of complementary experimental data such as excited-state lifetimes and branching ratios.
For many short-lived nuclei, especially on the neutron-rich side, precise information on the lifetimes of excited states is not known and thus another solution for the normalisation of the measured Coulomb-excitation cross sections needs to be applied.
In general two options are possible: either normalisation to the excitation of target nuclei with known electromagnetic matrix elements or to the number of elastically scattered beam particles.
 
This paper attempts to address some of the common problems and solutions encountered with the extraction of electromagnetic matrix elements from RIB Coulomb-excitation experiments in general, with examples taken from studies performed at REX-ISOLDE and GANIL.
Here, the \gosia{} code (see Section~\ref{sec:gosia}) is most commonly used for this purpose.
Firstly though, the observables from such experiments must be clearly defined; this is done in Section~\ref{sec:observables}.
Methods utilising the \gosia{} code for the analysis are presented in Section~\ref{sec:analysis} and a summary and outlook is given in Section~\ref{sec:summary}.

\section{The \gosia{} code} \label{sec:gosia}

Experiments performed in the 1950's utilising light-ion beams as a means of
exciting target nuclei were relatively simple to interpret using first- and
second-order perturbation theory.  Later, heavy-ion beam experiments
populated many excited states via multiple-step Coulomb excitation.  Early
versions of computer codes designed to handle the analysis of these data,
most notably that of Winther and de Boer~\cite{Alder1966}, employed the
semi-classical theory of multiple Coulomb excitation developed by Alder and
Winther~\cite{Alder1956}.  This code allowed quantitative calculations of
excitation amplitudes for the first time, using a set of reduced
electromagnetic matrix elements as input.  With this philosophy, the
\gosia{} code~\cite{Czosnyka1983} was designed in 1980 to achieve an
extraction of the electromagnetic matrix elements from a set of
Coulomb-excitation data by performing a fitting routine using these matrix
elements as parameters.  Both excitation and the consequent $\gamma$-ray
de-excitation, governed by the very same set of matrix elements, are
calculated within the code, allowing for a direct comparison to experimental
data~\cite{Czosnyka1983}.
The description of $\gamma$-ray de-excitation in \gosia{} is based on the \textsc{cegry} code~\cite{Cline1974,GosiaManual} and takes into account the angular correlations, deeorientation effect, recoil effects, Jacobian to a common reference frame, and integration over detector geometry. 

The first successful application of \gosia{} was to prove that the set of matrix elements obtained for $^{110}$Pd~\cite{Hasselgren1981} constituted a unique solution, which has been later confirmed by the results of a recoil-distance lifetime measurement~\cite{Kotlinski1984}.
The \gosia{} code was further validated and tested in the analysis of an extensive data set for several W-Os-Pt isotopes~\cite{Wu1983,Wu1989,Wu1991,Wu1996}.
In the following years, it was used to study shape evolution and coexistence in many regions of the nuclear chart, including transitional nuclei~\cite{Fahlander1988,Kotlinski1990Ge,Kavka1995}, rare earths~\cite{Kotlinski1990Er} and actinides~\cite{Czosnyka1986}, as well as exotic octupole shapes~\cite{Ibbotson1993}.


\section{Observables in Coulomb-excitation experiments} \label{sec:observables}

The direct observables in Coulomb-excitation experiments are usually the $\gamma$-ray intensities corresponding to the scattering of the projectile particle defined by the observation of at least one of the collision partners in a given angular and energy range.
In contrast, the deduced matrix elements are not direct observables and usually occur as strongly correlated parameters in a fit of the $\gamma$-ray intensity data.
In order to relate these gamma-ray intensities to the excitation cross sections of the populated states, which can be calculated for a given set of scattering and nuclear parameters, normalisation factors need to be introduced as described in Section~\ref{sec:analysis}.

\begin{figure}[tb]
\centering
\includegraphics[width=1.0\columnwidth]{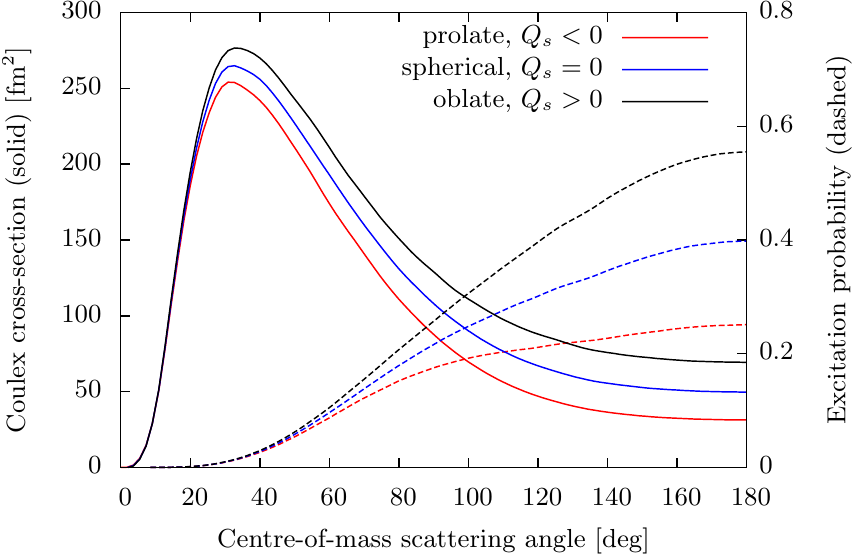}
\caption{Coulomb-excitation probabilities (dashed lines) and cross sections (solid lines; product of the Rutherford cross section and Coulomb excitation probability) for populating the $2^{+}_{1}$ state in $^{184}$Hg incident on a $^{120}$Sn at 2.8~MeV/$u$ under the different assumptions for the spectroscopic quadrupole moment, $Q_{s}$. The oblate (black) assumption is that of $Q_{s}=1.15$~$e$b, extracted from the measured $B(E2;2^{+}_{1}\rightarrow0^{+}_{1})$~\cite{Gaffney2014,Bree2014} and the rigid-rotor model ($K=0$), while the prolate assumption (red) has the same magnitude, but a negative
sign for $Q_{s}$. The spherical assumption ($Q_{s}$=0) is shown in blue.}
\label{fig:cexs}
\end{figure}

Data sets introduced to \gosia{} are most often described in terms of ``experiments''.
These may be defined by different combinations of beam and target, beam energy and scattering-angle range.
With the use of segmented particle detectors, such as the Double-Sided Silicon Strip Detectors (DSSSD) or Parallel Plate Avalanche Counters (PPAC), subdivision of the data can be made in terms of scattering angle, gaining sensitivity to second-order effects such as the spectroscopic quadrupole moment, $Q_{s}$. 
This is demonstrated in Figure~\ref{fig:cexs} where the reorientation
effect~\cite{Breit1956,deBoer1968} leads to an increasing deviation in the cross-section at large scattering angles for the assumption of different quadrupole moments.
This can be further increased by the use of different targets to disentangle contributions from single- and multiple-step excitation processes.

\begin{figure}[tb]
{
\centering
\includegraphics[width=1.0\columnwidth]{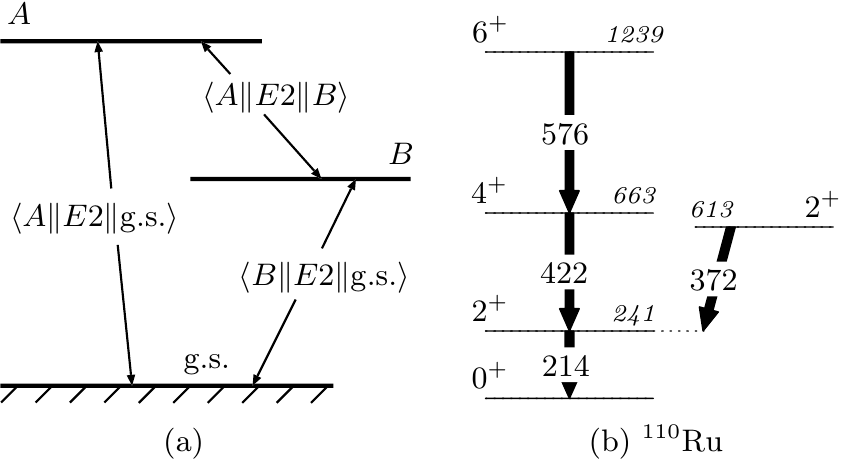}
\phantomsubcaption \label{fig:signs}
\phantomsubcaption \label{fig:110ru}
}
\caption{
(a)~Schematic level scheme showing the various excitation paths that lead to sensitivity to the relative signs of matrix elements, see text for details.
(b)~Low-lying states in $^{110}$Ru included in calculations shown in Fig.~\ref{fig:populations}.
}
\label{fig:levels}
\end{figure}

\begin{figure}[tb]
\centering
\includegraphics[width=1.0\columnwidth]{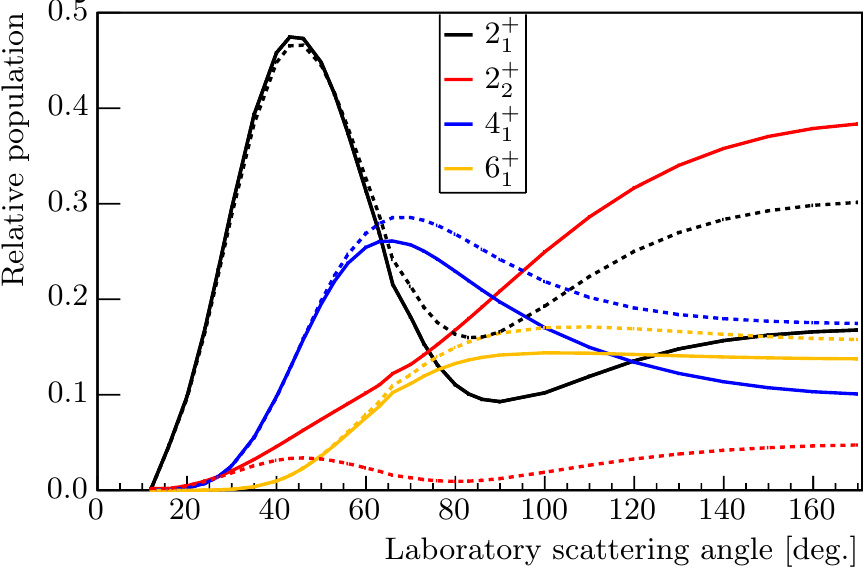}
\caption{Relative population of excited states in $^{110}$Ru Coulomb excited on $^{208}$Pb at 430~MeV beam energy, calculated for two different signs of $\langle 2^{+}_{1} \| E2 \| 2^{+}_{2} \rangle$: negative (solid lines) and positive (dotted lines), while all other matrix elements remain the same.
The $0^{+}_{1}$ ground state is not shown, but dominates the remainder of the population at all angles.} \label{fig:populations}
\end{figure}

It should be noted that in contrast to other spectroscopic methods Coulomb excitation is not only sensitive to magnitudes of the electromagnetic matrix elements, but also to their relative signs having a direct influence on excitation probabilities.
As an example one can consider a state $A$ that can be populated in one-step $E2$ excitation from the ground state or in a two-step $E2$ excitation process via an intermediate state $B$, as depicted in Figure~\ref{fig:signs}.
For each of the two possible excitation paths the contribution to the total excitation amplitude is proportional to the relevant matrix elements: $\langle A \| E2 \| \mathrm{g.s.} \rangle$ for the direct excitation and the product of $\langle A\| E2 \| B \rangle$ and \linebreak $\langle B \| E2 \| \mathrm{g.s.} \rangle$ for the two-step process.
The excitation \linebreak probability is proportional to the square of the sum of excitation amplitudes and therefore it contains not only quadratic terms ($\langle A \| E2 \| \mathrm{g.s.} \rangle^{2}$, related to $B(E2;A \to \mathrm{g.s.})$, and ${\langle A \| E2 \| B \rangle^{2} \langle B \| E2 \| \mathrm{g.s.} \rangle^{2}}$) but also interference \linebreak terms between possible excitation paths, such as \linebreak
${\langle A \| E2 \| \mathrm{g.s.} \rangle \langle A\| E2 \| B \rangle \langle B \| E2 \| \mathrm{g.s.} \rangle}$.
The signs of these interference terms depend on the relative signs of the matrix elements.
This is illustrated by the example of $^{110}$Ru [level scheme shown in Figure~\ref{fig:110ru}] on $^{208}$Pb presented in Fig.~\ref{fig:populations} where for large scattering angles the population of the 2$^{+}_{2}$ state depends very strongly on the sign of $\langle 2^{+}_{1} \| E2 \| 2^{+}_{2} \rangle$ with respect to those of $\langle 2^{+}_{1} \| E2 \| 0^{+}_{1} \rangle$ and  $\langle 2^{+}_{2} \| E2 \| 0^{+}_{1} \rangle$.
This effect can be strong enough to be visible even in low-statistics RIB measurements and thus for several exotic nuclei complete sets of matrix elements including their relative signs have been determined~\cite{Clement2007,Bree2014,Wrzosek-Lipska2015,Kesteloot2015}.  

\section{Coulomb-excitation data analysis} \label{sec:analysis}

\subsection{Normalisation of measured cross sections} \label{sec:analysis:normalisation}

In order to extract nuclear-structure parameters (matrix elements) from Coulomb-excitation data, the measured $\gamma$-ray intensities have to be converted to absolute excitation cross sections.
Possible complications arise from the fact that the efficiency of the particle detection set-up, dead-time, beam intensity etc. are not always known
with good precision.
To deal with this, \gosia{} uses normalisation constants, which relate the calculated and experimental intensities.
These can be fitted or given by the user, as described in the following sections.
In the most general form, the normalisation constant used in \gosia{} is a product of the Rutherford cross section, the time integrated beam current, the absolute efficiency of particle and $\gamma$-ray detection and the particle solid angle factor.
If the statistics are not sufficient to make use of particle-$\gamma$-ray angular correlations (which is usually the case for radioactive beam studies), $\gamma$-ray spectra from individual detectors may be summed together, reducing the number of necessary normalisation constants to one per experiment.
In such cases, the relative $\gamma$-ray detection efficiency as a function of energy has to be provided for each detector.

The normalisation constant, $C$, for a given experiment is fitted to all measured $\gamma$-ray intensities $I^{e}$ observed in an experiment by minimising the expression:
\begin{equation}
\sum_{i} (C I^{c}_{i} - I^{e}_{i})^{2} / \sigma^{2}_{i} 
\end{equation}
where $I^{c}_{i}$ denotes the calculated $\gamma$-ray intensity for the $i$-th observed transition integrated over beam energy and scattering angle, $I^{e}_{i}$ its measured intensity and $\sigma_{i}$ its experimental uncertainty.
Moreover, it is possible to introduce relative normalisation constants $C_{m}$ that link data sets resulting from the subdivision of data collected during one physical run into $m$ slices of scattering angle.
If for each of $m$ coupled experiments a relative normalisation constant $C_{m}$ is defined in \gosia{}, during the minimisation of the $\chi^{2}$ function the following expression is minimised and only one global normalisation constant $C_{\mathrm{global}}$ is fitted:
\begin{equation}
\sum_{m} \sum_{i} (C_{\mathrm{global}} C_{m} I^{c}_{i} - I^{e}_{i})^{2} / \sigma^{2}_{i}
\end{equation}
It should be noted that the $C_{m}$ factors can be arbitrarily rescaled, as the scaling factor can be always incorporated in $C_{\mathrm{global}}$.
The normalisation constants are fitted in \gosia{} at the same time as the matrix elements, during the minimisation of the $\chi^{2}$ function
described in Section~\ref{sec:analysis:chi2}.

The products $C_{m} I^{c}_{i}$ that are compared to experimental $\gamma$-ray intensities depend obviously both on the matrix elements and on the normalisation constants.
Especially in the case of one-step excitation, one can easily compensate a modification of the relevant matrix elements by adjusting the normalisation constant.
Therefore, in order to obtain a reliable set of matrix elements, additional constraints on either the matrix elements or the normalisation constants have to be provided. The possible techniques, depending on the specifics of the experiment, are presented in the following sections.

\subsubsection{Elastic scattering}

Historically, the simplest and most direct method of normalising Coulomb-excitation cross sections is to use the measured elastic-scattering (Rutherford) cross section. \linebreak
This requires precise knowledge of the scattering angular range, efficiency of the particle detection system and well understood dead time if one is to obtain the integrated beam current. Since the Rutherford cross section is very sensitive to scattering angle at low centre-of-mass angles, uncertainties related to geometry are minimised for backscattering as demonstrated in Ref.~\cite{Radford2002,Allmond2014}.
For inverse kinematics reactions, the backscattered projectiles are forward focused in the laboratory frame of reference and have low energy.
The corresponding recoils however, can be utilised where clean kinematic separation of these events can be made.
In RIB experiments, where beam intensities are low, the highest excitation probability is desired and as such, high-$Z$ targets are usually used.
Uncertainties are introduced because of events from different scattering angles that can be misinterpreted.
This is particularly true for many experiments utilising silicon strip detectors at forward angles, such as those at GANIL~\cite{Clement2007} and with Miniball at REX-ISOLDE~\cite{Warr2013}.

In cases where absolute Rutherford cross sections are not reliable (for example when downscaling is applied to single particle events and thus the dead time is different as compared to particle-gamma events), particle singles events may still be used to calculate the {\it relative} normalisation constants, $C_{m}$.
Commonly, this applies to experiments where data is taken in the same run but is divided into different angular cuts.
For this, one needs knowledge of the total number of scattered particles in each angular range, $N_{m}$, i.e. without a coincidence condition on $\gamma$~rays or a second particle.
The two are related by the following expression:
\begin{equation}
C_{m} = \frac{ N_{m} }{ \Delta\theta_{m} \Delta\phi_{m} }	,
\label{eq:normalisation:elastic}
\end{equation}
where ($\Delta\theta_{m},\Delta\phi_{m}$) represents the solid angle subtended in the experiment. Again, $C_{m}$ may be arbitrarily rescaled due to the remaining normalisation fitted by \gosia{}, $C_{\mathrm{global}}$, but the ratios of each coupled $C_{m}$ remains the same.

\subsubsection{Excited-state lifetimes or $B(E2)$ values}  \label{sec:analysis:normalisation:lifetime}

When multiple states are excited, with single- or multiple-step Coulomb excitation, one or more $B(E2)$ values connecting the ground-state and an excited state can be used to fit the normalisation constants for each experiment in \gosia{}, $C_{m}$.
For this, one must also observe the corresponding population of such a state with good precision, which means that the $\gamma$-ray intensity and efficiency, along with the branching ratio, must be known to good precision.
This is usually the simplest and preferred method in these cases as everything is fitted by the code and there are no additional calculations required by the user.

In even-even nuclei, the normalisation is usually fulfilled by an independent measurement of the $2^{+}_{1}$-state lifetime, $\tau(2^{+}_{1})$.
Two examples of this technique with RIBs, are the cases of $^{74,76}$Kr~\cite{Clement2007} and $^{182-188}$Hg~\cite{Bree2014,Bree2014-PhD,Wrzosek-Lipska2015}, where multiple lifetimes of yrast states were known in the literature and even re-measured~\cite{Gorgen2005,Grahn2009,Gaffney2014} to provide the required precision.
For odd-mass or odd-odd systems, multipole mixing ratios also become important since the strongest-observed $\gamma$~ray is often a mixed $E2/M1$ transition (see also Section~\ref{sec:analysis:sn}).
Furthermore, low-energy transitions in heavy nuclei can also be strongly converted, meaning that the strongest excitation path may not result in an intense $\gamma$-ray decay.
In these cases, it is usually possible to normalise to the next higher-lying transition since the low-energy of the first-excited state also means that the probability of two-step excitation approaches that of the single-step excitation, as was done in the analysis of $^{224}$Ra~\cite{Gaffney2013}.

\subsubsection{Target excitation} \label{sec:analysis:normalisation:target}
 
The electromagnetic interaction between the collision partners causes excitation of either the projectile or target nucleus. 
The observed excitation of target nuclei can usually be described with high precision using literature values of relevant matrix elements and used to normalise the excitation cross sections measured for beam nuclei.
The observed number of $\gamma$~rays in the transition de-exciting an excited state in the target nucleus, can be described in the following equation:
\begin{equation}
N_{t} = L \cdot \frac{\rho d N_{A}}{A_{t}} \cdot b_{t} \epsilon_{\gamma}(E_{t}) \epsilon_{\mathrm{part}} \sigma_{t} \label{eq:normalisation:target}
\end{equation}
where $\sigma_{t}$ is the integrated cross-section of exciting the given state in the target, $b_{t}$ is the total $\gamma$-ray branching ratio for the transition, $\epsilon_{\gamma}(E_{t})$ is the absolute efficiency of detecting a $\gamma$~ray of energy $E_{t}$, $\epsilon_{\mathrm{part}}$ is the efficiency of detecting a particle in the angular range defined by the integration limits of the cross-section, $\rho d$ is the thickness of the target in mg/cm$^{2}$, $N_{A}$ is Avogadro's number, $A_{t}$ is the mass number of the target and $L$ is the time-integrated luminosity of the beam. 
A similar equation can be constructed for the number of $\gamma$~rays in the transition de-exciting an excited state in the projectile, assuming the same angular range for particle detection: 
\begin{equation}
N_{p} =L \cdot \frac{\rho d N_{A}}{A_{t}} \cdot b_{p} \epsilon_{\gamma}(E_{p}) \epsilon_{\mathrm{part}} \sigma_{p}
\label{eq:normalisation:projectile}
\end{equation}

Taking a ratio of Equations \ref{eq:normalisation:projectile} and \ref{eq:normalisation:target} removes both the intrinsic particle detection efficiency and luminosity: 
\begin{equation}
\frac{N_{p}}{N_{t}} =
\frac{b_{p} \epsilon_{\gamma}(E_{p}) \sigma_{p}} {b_{t}\epsilon_{\gamma}(E_{t}) \sigma_{t}} 
\label{eq:normalisation:ratio}
\end{equation}
meaning that one can solve Equation~\ref{eq:normalisation:ratio} for $\sigma_p$ and there is no requirement to have knowledge of the integrated beam current. 
This is the principle of \gosia{}2.

When dealing with RIBs, pure beams are often not achievable and the target is also excited by beam contaminants.
If the beam composition is monitored during the experiment, this can be dealt with rather simply with the following correction to the experimental $\gamma$-ray intensities from the target~\cite{VandeWalle2009-Zn}:
\begin{equation} \label{eq:beamimpurities}
F = \frac{1}{ 1 + \sum_{c} \left( r_{c} \frac{\sigma_{t}\left(Z_{c},A\right)}{\sigma_{t}\left(Z_{X},A\right)} \right) }	,
\end{equation}
where $\sigma_{t}(Z,A)$ is the cross section of the target, excited by a beam with proton number $Z$ and mass $A$.
For every contaminant, $c$, with $Z=Z_{c}$, the ratio to the component of interest with $Z=Z_{X}$, can be expressed as $r_{c} = I_{c} / I_{X}$, where $I_{c,X}$ is the intensity of the respective components in the beam.

There also exists the possibility of impurities in the target. 
In this case the experimental intensities measured for the beam must be corrected to account for the scattering on target impurities.
For this, knowledge of the isotopic purity is required. 
This can be either from the target manufacturer or the observed excitation ratios, deduced from $\gamma$-ray intensities, if available.  Assuming only two components, a correction factor, $F_{i}$, can be calculated for each excited state, $i$~\cite{Kesteloot2015}:
\begin{equation}
\label{eq:targetimpurities}
F_{i} = \left( 1 + \frac{1}{P} \cdot \frac{\sigma_{i}(Z^{\prime},A^{\prime})}{\sigma_{i}(Z,A)} \right) , 
\end{equation}
where $\sigma_{i}(Z,A)$ and $\sigma_{i}(Z^{\prime},A^{\prime})$ are the excitation cross sections of a given state in the projectile on the main target species and contaminant, respectively. 
These can be calculated by \gosia{}, obtaining the ratio for each transition given a set of starting matrix elements.
The isotopic purity, $P$, is expressed by
\begin{equation}
\label{eq:targetpurity}
P = \frac{N_{A}}{N_{A^{\prime}}} , 
\end{equation}
where $N_{A,A^{\prime}}$ are the numbers of atoms of mass $A,A^{\prime}$.
By taking the ratio of the cross sections with different masses, at the same laboratory angles, the differences in Rutherford cross section and the centre-of-mass-dependent excitation probabilities are accounted for.
However, $F_{i}$ remains an estimation since the excitation probability of each state will depend in a complex manner on the electromagnetic matrix elements.
A systematic error must be retrospectively estimated due to this assumption by recalculating $F_{i}$ with the final set of matrix elements.
Differences between the original and final estimations of $F_{i}$ are usually small if $P$ is large.
In the case of $^{196}$Po on $^{94(95)}$Mo($P=95(2)\%$)~\cite{Kesteloot2015}, the maximum systematic error in $F_{i}$ was calculated to be 0.6\%, which is much smaller than the statistical uncertainty.

\subsection{$\chi^{2}$ square minimisation in \gosia{}} \label{sec:analysis:chi2}

The set of electromagnetic matrix elements is extracted by performing the minimisation of the $\chi^{2}$ function.
The total $\chi^{2}$ function is built of measured $\gamma$-ray intensities
and other known spectroscopic data, and those calculated from a set of matrix elements between all relevant
states.
The calculated $\gamma$-ray intensities are corrected for effects such as: internal conversion of electromagnetic transitions, the energy-dependent efficiency of the $\gamma$-ray detectors and the angular distribution of the emitted radiation.
A proper reproduction of the experimental $\gamma$-ray intensities requires integration over the scattering angular ranges, defined by the particle detection set-up, and over the range of incident projectile energies resulting from the energy loss in a target.
The convergence of the $\chi^{2}$ fit can be improved by using known spectroscopic data, e.g. $\gamma$-ray branching ratios, multipole mixing ratios or lifetimes.

The $\chi^{2}$ function consists of three components resulting from various subsets of
data:

\begin{equation} \label{eq:chisq}
\chi^{2} = S_{y} + S_{l} + S_{d}	 	.
\end{equation}

The first contribution, $S_{y}$, comes from the comparison of $\gamma$-ray intensities observed in the experiment, $I^{e}_{k}$, and those calculated with the fitted matrix elements, $I^{c}_{k}$, and is defined as:
\begin{equation} \label{eq:chisq-yields}
S_{y} = \sum_{ij} w_{ij} \sum_{k\left(ij\right)} \frac{1}{\sigma^{2}_{k}} \left( C_{ij} I^{c}_{k} - I^{e}_{k} \right)^{2} ,
\end{equation}
The summations extend over all defined experiments, $i$, $\gamma$-ray detectors, $j$, and the detector- and experiment- dependent number of observed transitions indicated by $k$.
The coefficients $C_{ij}$ are normalisation constants connecting experimental and calculated intensities.
These are equivalent to $C_{m}$ described in Section~\ref{sec:analysis:normalisation}, but the summation now extends independently over the number of independent $\gamma$-ray detectors as well as experiments or sub-divisions.
These are defined individually for each experiment and detector combination and fitted on the same basis as the matrix elements.
The weights, $w_{ij}$, ascribed to the various subsets of data defined by different experiments and $\gamma$-ray detectors, can be set independently by user.

The second contribution, $S_{l}$, is related with the user-defined ``observation limit'' and is defined as follows:
\begin{equation} \label{eq:chisq-limit}
S_{l} = \sum_{j} \left( \frac{I^{c}_{j}(i,j)}{I^{c}_{n}(i,j)} - u(i,j) \right)^{2}
\cdot \frac{1}{u^{2}(i,j)} .
\end{equation}
An experiment and detector dependent upper limit of $\gamma$-ray intensities, $u(i,j)$, is expressed as a fraction of the normalising transition specified by the user (usually it is the strongest observed transition, i.e., $2^{+}_{1}\rightarrow0^{+}_{1}$ for even-even nuclei).
If the calculated intensity of any unobserved $\gamma$-ray transition, divided by the intensity of the normalising transition, $I^{c}_{n}(i,j)$, exceeds this upper limit then it is included in the calculation of the least squares fit.
The summation extends over the calculated $\gamma$-ray transitions in each experiment and detector combination not defined as experimentally observed, provided that the upper limit has been exceeded.
A proper set of upper limits prevent finding unphysical solutions yielding the production of $\gamma$-ray transitions not observed in experiment.

The remaining term of Eq.~\ref{eq:chisq}, $S_{d}$, accounts for the additional spectroscopic data which can be included in the fit: lifetimes, branching ratios, multipole mixing ratios and known matrix elements. The summation extends over the number of such data points, $n_{d}$, given for each data type, $d$, and user-defined weights, $w_{d}$, which are common for a given group of spectroscopic data.
\begin{equation} \label{eq:chisq-specdata}
S_{d} =\sum_{d} w_{d} \sum_{n_{d}} \frac{1}{\sigma^{2}_{n_{d}}} \left( D^{c}_{n_{d}} - D^{e}_{n_{d}} \right)^{2} ,
\end{equation}
where $D^{c}_{n_{d}}$ and $D^{e}_{n_{d}}$ are the values of the spectroscopic data calculated using the current set of best-fit matrix elements and the experimental value, respectively.

A simultaneous fit of a large number of free parameters (matrix elements), having unknown correlations and possibly very different influences on the data, prevents precise determination of degrees of freedom.
In the simplest cases without these issues, the number of degrees of freedom would be defined as a result of the subtraction of the number of experimental data points and the number of fitted parameters.
The $\chi^{2}$ function resulting from the \gosia{} calculations is normalised to the number of data points, including experimental intensities, branching ratios, lifetimes, mixing ratios and known matrix elements.
In practical situations one deals exclusively with total $\chi^{2}$ values, thus the normalised $\chi^{2}$ value yielding from the \gosia{} code should be multiplied by the number of data points given, regardless of the user-defined weight,~$w$.

\subsubsection{The \gosia{}2 code} \label{sec:analysis:gosia2}

When lifetimes of the lowest excited states are not known with sufficient precision, the measured Coulomb-excitation cross-sections need to be normalised in a different way, for example to the target excitation, as described in Section~\ref{sec:analysis:normalisation:target}.
The \gosia{}2 code was developed to handle the simultaneous analysis of both target and projectile excitation.
The $\chi^{2}$ function of Eq.~\ref{eq:chisq} is minimised in parallel for the target and projectile whilst sharing the normalisation factors as parameters across both functions. 
Using literature values of relevant matrix elements in the target nucleus, the normalisation constants can be constrained by the $\gamma$-ray intensities of the target de-excitation.
The solution then corresponds to the global minimum of the total $\chi^{2}$ function defined as the sum of $\chi^{2}$ functions for both reaction partners.
If only two matrix elements are used to describe the excitation of the nucleus under study, a two-dimensional plot of the
total $\chi^{2}$ surface may be used to evaluate uncertainties of fitted matrix elements, as described in more detail in Section~\ref{sec:analysis:errors:2d}. 
However, there are certain limitations of the code: when more unknown matrix elements are involved, estimation of their errors becomes more complicated and one of the procedures described in Sections~\ref{sec:analysis:hg} and~\ref{sec:analysis:sn} are required.

\subsection{Methods of error estimation} \label{sec:analysis:errors}

\subsubsection{Standard error estimation in  \gosia{}}  \label{sec:analysis:errors:gosia1}

Statistical errors of the matrix elements are estimated after the convergence of the global minimum of the $\chi^{2}$ function and can be obtained from the probability distribution around the minimum. 
The applied method involves two steps.
At first, the diagonal, or uncorrelated, uncertainties are calculated by sampling each matrix element about the minimum of the $\chi^{2}$ surface, finding the point where an increase in $\chi^{2}$ is achieved, satisfying the $1\sigma$ condition.
This condition is defined by requesting that the total integrated probability distribution in the space of the fitted parameters be equal to the $1\sigma$ confidence limit~--~68.27~\%~\cite{Czosnyka1983}.
At the same time, a multi-dimensional correlation matrix is built, which is then used in the second step in order to compute the fully correlated errors on each matrix element, satisfying the same condition.

\subsubsection{Two-dimensional $\chi^{2}$ surface analysis} \label{sec:analysis:errors:2d}

\begin{figure}[tb]
{
\centering
\includegraphics[width=0.96\columnwidth]{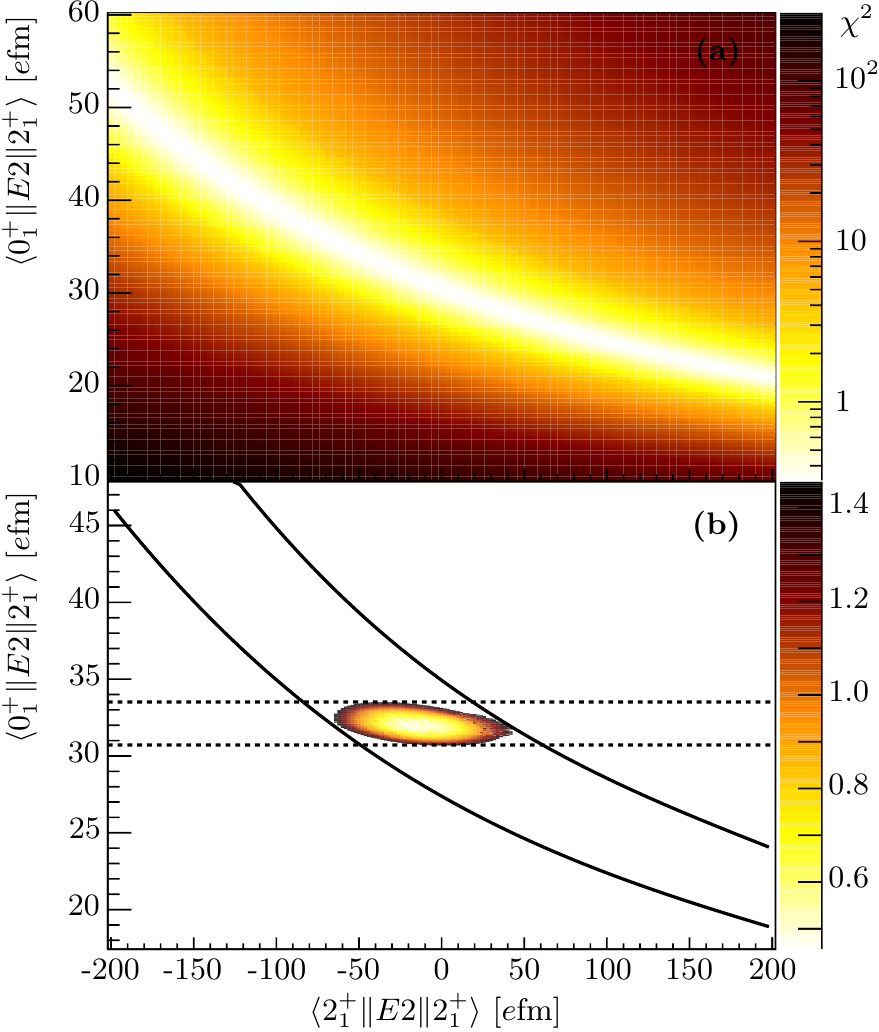}
\phantomsubcaption \label{fig:62Fe-a}
\phantomsubcaption \label{fig:62Fe-b}
}
\caption{
(a) A full two-dimensional $\chi^{2}$ surface with respect to $\langle 2^{+}_{1} \|E2\| 0^{+}_{1} \rangle$ and $\langle 2^{+}_{1} \|E2\| 2^{+}_{1} \rangle$ for the $^{62}$Fe projectile, reproduced from Ref.~\cite{Gaffney2015-MnFe}. The data are normalised to the excitation of a 4.0-mg/cm$^{2}$ thick $^{109}$Ag target at a beam energy of 2.86~MeV/$u$ using \gosia{}2.
(b) The resulting surface when combined with lifetime measurements~\cite{Ljungvall2010,Rother2011} and a $1\sigma$ cut applied with the condition that {$\chi^{2}<\chi^{2}_{\mathrm{min}}+1$}.
The individual $1\sigma$ contours for the Coulomb-excitation and lifetime data are shown by the solid and dashed lines, respectively.
Reproduced from Ref.~\cite{Gaffney2015-MnFe}}
\label{fig:62Fe}
\end{figure}

In a multi-parameter analysis, the global best fit can be found by
constructing a $\chi^{2}$ hyper-surface with respect to all parameters.  In
the case of a two-parameter system one is able to visualise a 2-dimensional
$\chi^{2}$ surface as shown Figure~\ref{fig:62Fe-a}.  Here the example is of
the two matrix elements, $\langle 2^{+}_{1} \|E2\| 0^{+}_{1} \rangle$ and
$\langle 2^{+}_{1} \|E2\| 2^{+}_{1} \rangle$, usually sufficient to describe
the excitation process of an even-even nucleus if only the $2^{+}_{1} \to
0^{+}_{1}$ transition is observed.  The minimum of such a surface,
$\chi^{2}_{\mathrm{min}}$, can easily be found and the 1$\sigma$-uncertainty
contour can be defined as the region of the surface for which
$\chi^{2}<\chi^{2}_{\mathrm{min}}+1$~\cite{Cline1970,Lesser1972,Rogers1975}.  This technique was
used for the analysis of $^{94,96}$Kr~\cite{Albers2012,*Albers2013}.

\begin{figure}[tb]
\centering
\includegraphics[width=1.0\columnwidth]{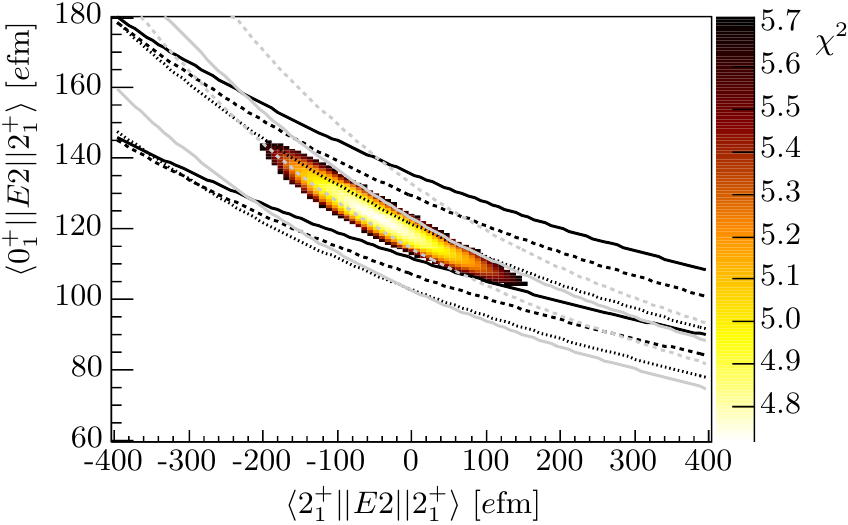}
\caption{
A two-dimensional $\chi^{2}$ surface with respect to $\langle 2^{+}_{1} \|E2\| 0^{+}_{1} \rangle$ and $\langle 2^{+}_{1} \|E2\| 2^{+}_{1} \rangle$ for $^{202}$Rn~\cite{Gaffney2015-Rn}. A $1\sigma$ cut is applied with the condition that $\chi^{2}<\chi^{2}_{\mathrm{min}}+1$ . The data is normalised to the excitation of a 4.0-mg/cm$^{2}$ thick $^{109}$Ag target at a beam energy of 2.90~MeV/$u$ using \gosia{}2. The data was sub-divided into five different scattering angular ranges and their individual $1\sigma$ limits are represented by the different bands; in increasing order of centre-of-mass scattering angle these are: solid black, dashed black, dotted black, solid grey and dashed gray.}
\label{fig:204Rn}
\end{figure}

If one of the parameters is independently measured, e.g. via lifetime measurements, the $\chi^{2}$ surface can be easily recalculated by adding the $\chi^{2}$ contribution of the new measurement at every point.
A new 1$\sigma$ contour is then also defined, as shown in Figure~\ref{fig:62Fe-b}.
This goes too for other independent Coulomb excitation measurements, which \linebreak may come from the segmentation of a data set into angular ranges (see example in Figure~\ref{fig:204Rn}) or different targets as described earlier.
The final uncertainties are obtained by projecting the $1\sigma$ uncertainty contour on the respective axes.
While the projected uncertainties are useful for understanding the precision on a given spectroscopic observable, such as $B(E2)$ values or spectroscopic quadrupole moments, the existing correlation between these parameters is lost.

In the past, the assumption that the influence of the spectroscopic quadrupole moment, $Q_{s}$, is negligible, or that otherwise its value can be assumed to be equal to zero has sometimes been used.
In the case of $^{62}$Fe, if it was not for the independent lifetime experiments, shown by the solid black line in Figure~\ref{fig:62Fe-b}, the 2-dimensional $1\sigma$ surface would not be constrained.
One possibility is to project such a surface with a single value of $Q_{s}$, or calculate only a 1-dimensional surface at a fixed value of $Q_{s}$.
However, this would greatly underestimate the uncertainty, since correlations are ignored.
In the $^{204}$Rn example of Figure~\ref{fig:204Rn}, this leads to a factor of 3.5 reduction in the true uncertainty.
Instead, it would be preferable in these cases to use a model assumption where necessary to provide limits of $Q_{s}$ as a function of $\langle 2^{+}_{1} \|E2\| 0^{+}_{1} \rangle$, for example the rigid rotor model~\cite{BohrMottelson}.
The total surface can then still be constrained but with a reasonable consideration of the uncertainty due to the influence of $Q_{s}$.

The graphical method however, becomes computationally time consuming and visually useless as the number of parameters increases.
Therefore alternative solutions of error estimation are proposed and some examples are presented in the following sections.
Their applicability depends on the strength of the correlations between matrix elements.

\subsection{Selected applications}
\subsubsection{Normalisation to the $B(E2)$ extracted from data sets where no correlations are observed} \label{sec:analysis:44ar}

The influence of the quadrupole moment of a given state on its excitation probability varies significantly with scattering angle as shown in Figure~\ref{fig:cexs}.
This dependence can be exploited in order to determine both the transition probability and the diagonal matrix element, even if only one $\gamma$-ray transition is observed in the nucleus of interest.
If the particle detector covers a sufficiently broad range of centre-of-mass scattering angles, the simplest solution described in Ref.~\cite{Zielinska2009} can be applied.
Here, in the first step, the ${B(E2;2^{+}_{1} \to 0^{+})}$ value is derived from the excitation cross-section of the $2^{+}_{1}$ state for the lowest scattering angles.
The influence of the quadrupole moment, $Q_{s}(2^{+}_{1})$, on the excitation probability of the $2^{+}_{1}$ state for this range of scattering angles was estimated at $4\%$ , which was below the statistical error of $7\%$ of the corresponding $\gamma$-ray intensity.
It was therefore a reasonable approximation to assume that in this case the observed transition strength depends only on the transitional matrix element.
The adopted uncertainty of the $B(E2;2^{+}_{1} \to 0^{+}_{1})$ included contributions from the statistical error of measured $\gamma$-ray intensities in $^{44}$Ar and $^{109}$Ag, as well as the uncertainty on the relative $\gamma$-ray efficiency, target matrix elements and the systematic error of 4\% resulting from neglecting $Q_{s}(2^{+}_{1})$ in the Coulomb-excitation calculations.
In the second step, this $B(E2;2^{+}_{1} \to 0^{+}_{1})$ and its uncertainty were used in the further analysis as an additional data point in a \gosia{} fit.
The remaining data was then subdivided into three angular ranges, with the width and number of ranges being chosen to obtain the maximum sensitivity to $Q_{s}(2^{+}_{1})$.
The $\gamma$-ray intensities of $^{44}$Ar from these ranges were normalised to the intensity measured for the first range, with relative normalisation factors fitted using the corresponding $^{109}$Ag $\gamma$-ray intensities.
Then the standard version of the \gosia{} code was used to simultaneously fit all the transitional and diagonal matrix elements to the measured intensities.

\subsubsection{Multiple Coulomb excitation and normalisation with a dominant transition to target excitation: combined \gosia{}-\gosia{}2 analysis} \label{sec:analysis:hg}

In multiple Coulomb excitation of even-even nuclei, several states can be populated. In such cases the $2^{+}_{1}$ state is usually dominantly populated as compared to other excited states.
When the lifetime of the $2^{+}_{1}$ state is not known with sufficient precision and the $B(E2;2^{+}_{1}\rightarrow 0^{+}_{1})$ value cannot be extracted as described in Section~\ref{sec:analysis:44ar}, measured Coulomb-excitation cross sections need to be normalised in a different way using e.g., target excitation.
However, a full analysis with the \gosia{}2 code, as presented in Section~\ref{sec:analysis:sn}, is not possible as the number of parameters increases significantly. The error estimation including correlations between all matrix elements involved becomes very complex and practically impossible.
A different solution needs to be found that handles both aspects: (i) normalisation to the target excitation and, (ii) error calculations including correlations between all matrix elements.
In such cases a combined analysis is required with the use of both standard \gosia{} and \gosia{}2 codes.

In the first step, a simplified analysis is performed aiming to determine the $B(E2;2^{+}_{1}\rightarrow 0^{+}_{1})$ value for the projectile.
Only one-step excitation of the $2^{+}_{1}$ state is considered, taking into account that population of the $2^{+}_{1}$ state depends predominantly on both the $B(E2;2^{+}_{1}\rightarrow 0^{+}_{1})$ value and spectroscopic quadrupole moment, $Q_{s}(2^{+}_{1})$.
In order to gain sensitivity on the extraction of the quadrupole moment of the $2^{+}_{1}$ state, the data are divided in terms of particle-scattering angular range.
The influence of the multi-step excitations resulting in population of higher-lying states is not usually included at this stage, although the level energies and a set of fixed ``starting'' matrix elements can be declared if
reasonable assumptions can be made concerning their magnitudes and relative
signs.
The analysis is performed as described in Section~\ref{sec:analysis:gosia2} using the \gosia{}2 code.
As a result a two-dimensional total $\chi^{2}$ surface (being the sum of
$\chi^{2}$ for the projectile and target system) as a function of the $B(E2;2^{+}_{1}\rightarrow 0^{+}_{1})$ value and the quadrupole moment $Q_{s}(2^{+}_{1})$ is determined and reflects correlations between these two parameters.
The final values are determined by the minimum of the $\chi^{2}$ function and their error bars are obtained by projecting the $1\sigma$-contour on the respective axes, as in Section~\ref{sec:analysis:errors:2d}.
The extracted $B(E2;2^{+}_{1}\rightarrow 0^{+}_{1})$ value is a first approximation. Its uncertainty includes: (i) the uncertainties of the $\gamma$-ray intensities originating from the target excitation, (ii) the uncertainties of the $\gamma$-ray intensities originating from the projectile excitation and, (iii) the uncertainties of the relevant, literature $B(E2)$ values for the target nucleus.

In the second step, correlations with all remaining matrix elements, which couple higher-lying excited states observed in the experiment have to be investigated.
This is performed using the standard \gosia{} code with full error estimation procedure (see Section~\ref{sec:analysis:errors:gosia1}) implemented in \gosia{}.
All states populated in the Coulomb-excitation experiment, together with all observed $\gamma$-ray intensities are taken into account in this part of analysis.
All involved electromagnetic matrix elements are now introduced as well.
Data extracted from the simplified \gosia{}2 analysis, specifically $\langle 0^{+}_{1} \| E2 \| 2^{+}_{1} \rangle$, serves as an absolute normalisation for the standard \gosia{} calculations.
It is declared together with its uncertainty as an additional data point and thus it is treated in the fit on equal rights as the $\gamma$-ray intensities.
Other spectroscopic data i.e., $\gamma$-ray branching ratios, mixing ratios, can also be included at this stage of analysis if known.
Note that the $\langle 2^+_1 \| E2 \| 2^+_1 \rangle$ diagonal matrix element extracted for the projectile in the first part of the analysis is not included as an additional data point in the fit when switching to the standard \gosia{} calculations.
Information on $\langle 2^+_1 \| E2 \| 2^+_1 \rangle$ is implicitly given by the relative normalisation constants extracted from the target excitation linking different angular data subdivisions.

In order to link each data set resulting from subdivision into several particle-scattering angular ranges, the relative normalisation constants are required.
These are usually calculated from the target excitation. The standard \gosia{} fit of observed $\gamma$-ray intensities depopulating excited states in the target nucleus is performed using literature values of all relevant matrix elements (see Section~\ref{sec:analysis:normalisation:lifetime}).
Calculated relative normalisation constants, $C_{ij}$, for each data set are then further used to fit the projectile excitation.
A small correction is applied here to achieve the same relative normalisation constants obtained in the \gosia{}2 solution, where the projectile data is also considered.
This is calculated using the ratio of the calculated yields for the normalisation transition in the target, $I^{c}_{n}(i,j)$, from the \gosia{}2 and standard \gosia{} solutions.
The fixed coupling of the relative normalisation constants removes the knowledge of the uncertainty in their ratio. In order to preserve such information in the fit, it is included indirectly.
The uncertainty of the $\gamma$-ray intensities related to the $2^{+}_{1} \to 0^{+}_{1}$ transition in the projectile, $\Delta I^{p}_{\gamma}(2^{+}_{1} \to 0^{+}_{1})$, to which we are normalising, is defined in \gosia{} so that it encompasses the uncertainty from the target excitation:
\begin{equation} \label{eq:normalisation-uncertainty}
{\Delta I^{p}_{\gamma}}^{2} = 
{\Delta^{\prime} I^{p}_{\gamma}}^{2}  +
 {I^{p}_{\gamma}}^{2} \cdot \left(  \sum_{i} \frac{1} { {\Delta I^{t}_{\gamma_{i}}}^{2} } \right)^{-1} 	,
\end{equation}
where $I^{p}_{\gamma}$ and $\Delta^{\prime} I^{p}_{\gamma}$ are the efficiency-corrected intensity and its associated uncertainty, of the $2^{+}_{1} \to 0^{+}_{1}$ transition in the projectile, respectively, and $\Delta I^{t}_{\gamma_{i}}$ can be expressed as:
\begin{equation} \label{eq:normalisation-uncertainty-target}
{\Delta I^{t}_{\gamma_{i}}}^{2} = 
\left( \frac{ \Delta^{\prime} I^{t}_{\gamma_{i}} }{ I^{t}_{\gamma_{i}} } \right)^{2}  + 
\left( \frac{ \Delta B(E2; i \to g.s.) }{ B(E2; i \to g.s.) }  \right)^{2}		,
\end{equation}
where $I^{t}_{\gamma_{i}}$ and $\Delta^{\prime} I^{t}_{\gamma_{i}}$ are the sum of efficiency-corrected intensities and associated uncertainties (in quadrature) of transitions depopulating a state $i$ in the target, respectively.
This assumes that this state is dominated by single-step excitation from the ground state and consequently by $B(E2; i \to g.s.)$ and its uncertainty, $\Delta B(E2; i \to g.s.)$.

\begin{figure}[tb]
\centering
\includegraphics[width=1.0\columnwidth]{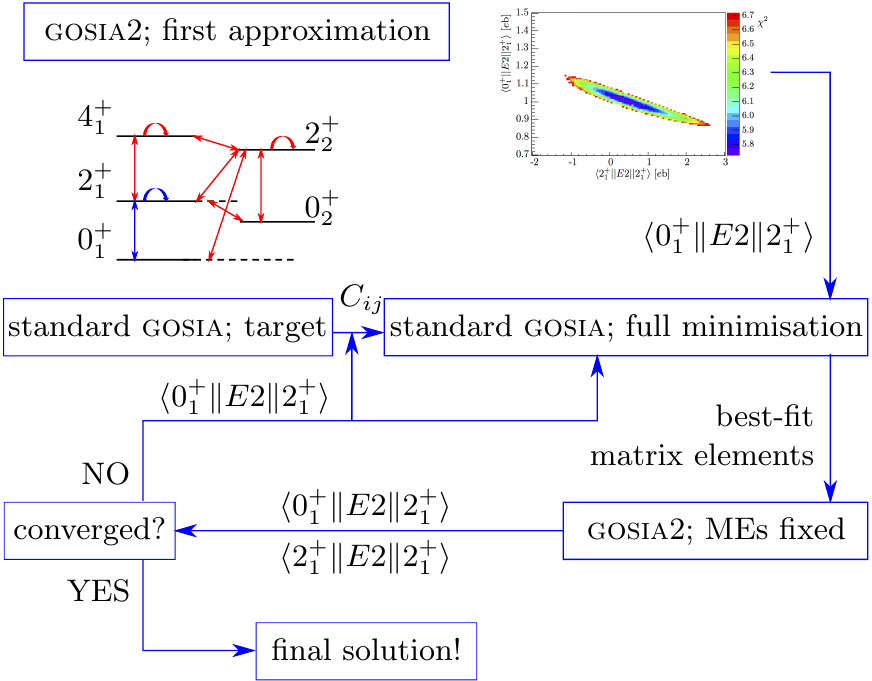}
\caption{A scheme of the combined analysis performed with the standard
\gosia{} and \gosia{}2 codes.  The presented method is used when
normalisation to the target excitation is required in multiple-step Coulomb
excitation of even-even nuclei.  The red matrix elements in the level scheme
of the figure are kept fixed during the \gosia{}2 calculations, while the
blue matrix elements are scanned to produce a 2-dimensional total $\chi^{2}$
surface plot (top right).  All matrix elements are varied in the full
\gosia{} minimisation and the best fit values are used in the next \gosia{}2
calculation.  Convergence is reached when the blue matrix elements are
consistent in both \gosia{} and \gosia{}2 calculations.}
\label{fig:flowchart}
\end{figure}

As a result of the second part of the analysis with the use of the standard \gosia{} code, a set of electromagnetic matrix elements between all states populated in the experiment is extracted.
Note that the $\langle 0^{+}_{1} \| E2 \| 2^{+}_{1} \rangle$ matrix element, used as an absolute normalisation for the full standard \gosia{} fit, originates from the simplified \gosia{}2 calculations where multiple Coulomb excitation was not
necessarily correctly considered. This influence needs to be taken into account.
For this purpose, the \gosia{}2 calculations have to be repeated using the set of matrix elements extracted in the second step of the analysis.
Only $\langle 0^{+}_{1} ||E2|| 2^{+}_{1} \rangle$ and $\langle 2^{+}_{1} \|
E2 \| 2^{+}_{1} \rangle$ for the projectile are \linebreak scanned as in the
first approximation, while all the other matrix elements for the projectile
are fixed and those for the target remain free.  As a result, a new total
$\chi^{2}$ surface is calculated.
Again, the $\langle 0^{+}_{1} \| E2 \| 2^{+}_{1} \rangle$ matrix element is determined from the $\chi^{2} < \chi^{2}_{\mathrm{min}} + 1$ condition.
It may differ from the value obtained from the first approximation since the correlations with other matrix elements will be different.
If this is the case, a full standard \gosia{} analysis with the updated value of the $\langle 0^{+}_{1} \| E2 \| 2^{+}_{1} \rangle$ matrix element has to be repeated in order to achieve consistency.
The whole standard \gosia{}~--~\gosia{}2 procedure should be iterated until the converged solution for both transitional and diagonal matrix
elements for the 2$^{+}_{1}$ state is obtained.
A schematic procedure of the \gosia{}~--~\gosia{}2 analysis is presented in Figure~\ref{fig:flowchart}.

In some cases, such as $^{196}$Po~\cite{Kesteloot2015}, the $\gamma$-ray intensity of higher-lying transitions is too weak to be reliably observed in each of the angular subdivisions. For this data to be included, an additional data set must be declared in the \gosia{} stage of the analysis that represents the sum total of all angular ranges.
The simplest way to normalise this data is to use the total intensity of the $2^{+}_{1} \to 0^{+}_{1}$ normalisation transition allied with the $B(E2)$ value that is already declared.
This intensity then exclusively constrains the absolute normalisation of the total data set with an uncertainty determined by the combination of the $B(E2)$ and $I_{\gamma}(2^{+}_{1} \to 0^{+}_{1})$ uncertainties.
During the correlated error calculation, this uncertainty on the absolute normalisation is effectively propagated to the higher-lying transitions.

\subsubsection{Normalisation to target excitation when multiple single-step excitations are observed}  \label{sec:analysis:sn}

When multiple, single-step excitations are observed with similar intensity, such as in odd-mass systems, there are too many parameters to make
an analysis of a  full $\chi^{2}$ hyper-surface feasible.
Instead, a one-dimensional surface is constructed for each matrix element by scanning the parameter to be investigated. At each point, the investigated parameter is kept fixed while all others are minimised with respect to $\chi^{2}$.
For this, the minimisation procedure of \gosia{}2, described in Section~\ref{sec:analysis:gosia2}, is invoked.
This procedure traces the lowest value path through the valley of the hyper-surface, effectively projecting the correlated surface to a given parameter.
The constructed surface can then be used in order to extract the $1\sigma$ uncertainty using the standard $\chi^{2}_\mathrm{min}+1$ method~\cite{Rogers1975}.
There is an assumption here of parabolic behaviour about the minimum, which for strongly correlated systems may not necessarily be true and asymmetric limits may be obtained.

Computationally, the time involved to minimise the full parameter space hundreds of times is very large.
For this reason, alternative methods of normalisation are preferred, but Coulomb excitation of odd-mass or odd-odd systems with RIBs tend to lack the required lifetime and multipole mixing ratio data to sufficient precision. 
This approach has been successfully used for the analysis of Miniball experiments on odd-mass Sn isotopes~\cite{DiJulio2012,*DiJulio2012a} and the odd-odd $^{26}$Na~\cite{Siebeck2015}.

\subsubsection{Normalisation to target excitation in a strongly correlated odd-mass system}
\label{sec:97rb}

\begin{figure}[tb]
\centering
\includegraphics[width=1.0\columnwidth]{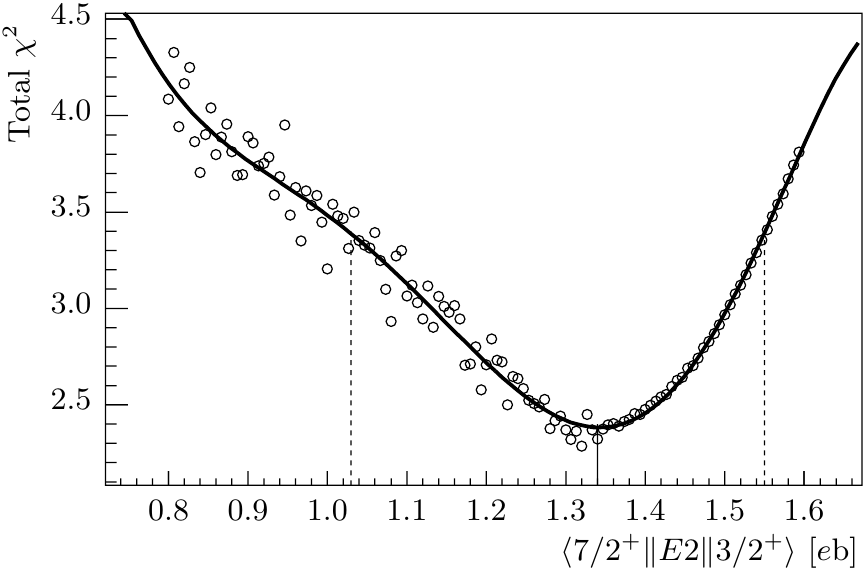}
\caption{
Total $\chi^{2}$ as a function of $\langle 7/2^{+} \| E2 \| 3/2^{+} \rangle$ in $^{97}$Rb. 
The open points show the $\chi^{2}$ obtained after convergence of the minimisation procedure, and the solid line is a polynomial fit of the $\chi^{2}$ distribution.
The vertical lines correspond to $\chi^{2}_{\mathrm{min}}$ (solid) and $\chi^{2} = \chi^{2}_{\mathrm{min}}+1$ (dashed; 1$\sigma$ error bar).  }
\label{figure_chi2}
\end{figure}

In the example of $^{97}$Rb, a strongly-deformed band built on the $3/2^{+}$ ground state is populated in Coulomb excitation with a $^{60}$Ni target~\cite{Sotty2015}. 
Mixed $E2/M1$ $I \to I-1$ transitions are roughly one order of magnitude stronger in intensity than $I \to I-2$ transitions.
In order to extract transition probabilities in the low-energy part of the band, normalisation to target excitation is necessary.
On the other hand, transition probabilities between the states that can only
be reached in multi-step excitation are related to measured intensity ratios
in the nucleus of interest.
As an example, the $4^{+}_{1} \to 2^{+}_{1}$/$2^{+}_{1} \to 0^{+}_{1}$ intensity ratio observed in Coulomb excitation of a weakly deformed even-even nucleus, assuming quadrupole moments equal to zero, depends exclusively on the $\langle 4^{+}_{1} \| E2 \| 2^{+}_{1} \rangle$ matrix element; changing the $\langle 0^{+}_{1} \| E2 \| 2^{+}_{1} \rangle$ matrix element would influence the total number of counts in both transitions, but not the ratio.
This is no longer true if a significant fraction of nuclei (few percent) undergo excitation in each step of the process, which is often the case of deformed nuclei including $^{97}$Rb, but still observed relative intensities in the upper part of the band depend only weakly on lifetimes of the lowest-excited states.
Therefore the analysis can be divided in two parts: the $\langle 7/2^{+} \| E2 \| 3/2^{+} \rangle$ matrix element is determined using normalisation to target excitation (\gosia{}2 analysis), and the remaining matrix elements are extracted from the intensities measured for $^{97}$Rb using the \gosia{} code, fixing $\langle 7/2^{+} \| E2 \| 3/2^{+} \rangle$ at the value determined in the first part of the analysis.
The choice of this matrix element was due to the fact that it corresponds to the only pure $E2$ transition from the ground state.
The \gosia{}2 code is used to find a minimum of the $\chi^{2}$ function resulting from comparison of measured and calculated $\gamma$-ray intensities in $^{97}$Rb and $^{60}$Ni, as well as known spectroscopic data in $^{60}$Ni ($B(E2; 2^{+} \to 0^{+}$) and $Q_{s}(2^{+}_{1})$).
Measured intensities of the $2^{+} \to 0^{+}$ transition in $^{60}$Ni were scaled according to the measured beam composition, and their statistical uncertainties were adjusted to take into account the uncertainty of the beam composition, as described in Section~\ref{sec:analysis:normalisation:target}. 
The minimisation is performed for several hundred starting values of $\langle 7/2^{+} \| E2 \| 3/2^{+} \rangle$ ranging from 0 to 3 eb. 
The $\langle 7/2^{+} \| E2 \| 3/2^{+} \rangle$ matrix element was fixed during the minimisation, while all other matrix elements are allowed to vary, with only constraints resulting from Alaga rules~\cite{Alaga}. 
In this way correlations between matrix elements are taken into account.
Figure~\ref{figure_chi2} presents the $\chi^{2}$ distribution as a function of $\langle 7/2^{+} \| E2 \| 3/2^{+} \rangle$ in the vicinity of minimum.
The vertical lines correspond to the adopted mean value (minimum of the $\chi^{2}$ distribution) and error bars ($\chi^{2}=\chi^{2}_{\mathrm{min}} +1$) for the $\langle 7/2^{+} \| E2 \| 3/2^{+} \rangle$ matrix element.

The second part of the analysis is performed using \gosia{} with the $\langle 7/2^{+} \| E2 \| 3/2^{+} \rangle$ matrix element fixed at the value determined in the first part of the analysis.
The errors of all remaining matrix elements are estimated using the standard error evaluation procedure implemented in \gosia{} (see Section~\ref{sec:analysis:errors:gosia1}). 
For transitions de-exciting states up to $11/2^{+}$, it also is necessary to propagate the uncertainty of $\langle 7/2^{+} \| E2 \| 3/2^{+} \rangle$.
For higher-lying transitions, contributions of this source of error to the total uncertainty is determined to be negligible.
In this part of the analysis, $\langle 7/2^{+} \| E2 \| 3/2^{+} \rangle$ is fixed instead of being fitted as an additional data point in order to make sure that  its uncertainty is properly propagated.
When $\langle 7/2^{+} \| E2 \| 3/2^{+} \rangle$, with the uncertainty determined in the first part of the analysis, is simply included in the fit on an equal basis as the $\gamma$-ray transition intensities, its final uncertainty (and in consequence those of other matrix
elements) is underestimated by the standard procedure of errors evaluation in \gosia{}, as the $\chi^{2}$ minimum with respect to this matrix element is artificially made deeper by including two data points corresponding to the same observable ($\langle 7/2^{+} \| E2 \|3/2^{+} \rangle$ and the $7/2^{+} \to 3/2^{+}$ transition intensities) in the fit.
Such an effect has not been observed in the combined \gosia{}-\gosia{}2 analysis  of Coulomb-excitation data in even-even nuclei (see \linebreak Section~\ref{sec:analysis:hg}) since there the $2^{+}_{1} \to g.s.$ transition is dominant, known with better precision than other $\gamma$-ray intensities and thus its intensity serves basically to calculate the relative normalisation parameters for each experiment.
In the case of $^{97}$Rb, the $7/2^{+} \to 3/2^{+}$ transition is roughly 20 times weaker than the strongest $I \to I-1$ transitions in this nucleus, which are consequently used to calculate the relative normalisation parameters.

\subsubsection{Normalisation to transition intensities in the nucleus of interest}

\begin{figure}[tb] 
\centering
\includegraphics[width=1.0\columnwidth]{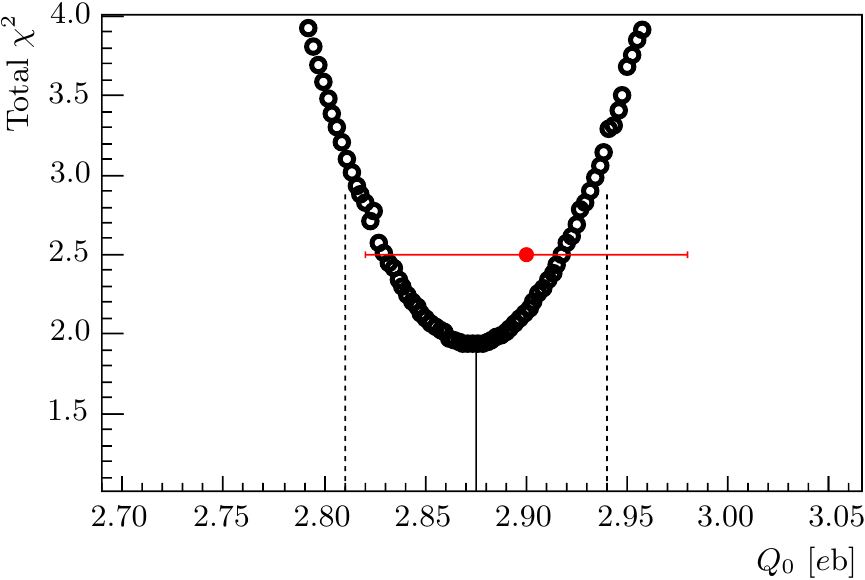}
\caption{
Total $\chi^{2}$ as a function of $Q_0$ in $^{97}$Rb under assumption of the rigid rotor model for all $E2$ matrix elements.
The open points show the $\chi^{2}$ obtained after convergence of the minimisation procedure, the vertical lines correspond to $\chi^{2}_{\mathrm{min}}$ (solid) and $\chi^{2} = \chi^{2}_{\mathrm{min}}+1$ (dashed; 1$\sigma$ error bar).
The weighted average of $Q_0$ values calculated from individual $E2$ matrix elements obtained in the full Coulex analysis including normalisation to the target excitation is shown in red.
}
\label{fig:chi2_Q0}
\end{figure}

In very favourable cases of collective nuclei an estimation of transition probabilities can be obtained from the ratios of transition intensities in the nucleus of interest.
It requires, however, strong model assumptions concerning the collectivity of the states (purely rotational or vibrational character).
This procedure has been tested on the  $^{97}$Rb data where all $E2$ matrix elements between the observed states were coupled assuming the rigid rotor model.
In this way one single parameter, corresponding to the transitional quadrupole moment $Q_0$ of the band, was used to describe the $E2$ part of the measured gamma-ray intensities.
No assumptions were made on the $M1$ matrix elements of the mixed $E2/M1$ $I \to I-1$ transitions and in total 7 parameters (one $Q_0$ value and 6 M1 matrix elements) were fitted to twenty measured $\gamma$-ray intensities.
In order to estimate the uncertainty of the extracted $Q_0$ value, the minimisation procedure was performed again for several hundred values of $Q_0$ kept fixed during minimisation with $M1$ matrix elements free to vary. 
A distinct minimum of the  $\chi^{2}$ distribution was found as shown in Figure~\ref{fig:chi2_Q0}.
Both the obtained value of $Q_0$, as well as the error bars corresponding to $\chi^{2}=\chi^{2}_{\mathrm{min}} +1$ are consistent with the weighted average of $Q_0$ values calculated from individual $E2$ matrix elements obtained in the full Coulomb-excitation analysis including normalisation to the target excitation, presented in Sec.~\ref{sec:97rb}.

\subsection{Dealing with non-standard particle detectors}

Particle detectors used for RIB Coulomb-excitation experiments are usually axially symmetrical and have an absolute efficiency close to 100\%.
As long as the efficiency is uniform, any deviations from 100\% are included in the normalisation constants (see Section~\ref{sec:analysis:normalisation}).
However, with the expected increase of RIB intensities, the standard annular Si detectors that are currently used will likely be replaced by more complicated particle detection set-ups, possibly consisting of various types of detectors differing in efficiency.
In addition, radiation damage may deteriorate parts of a detector, resulting in a very complicated shape in the $\theta$-$\phi$ plane.

\subsubsection{Complex particle-detector shapes} \label{sec:detectorshapes}

\begin{figure}[tb]
\centering
\includegraphics[width=0.98\columnwidth]{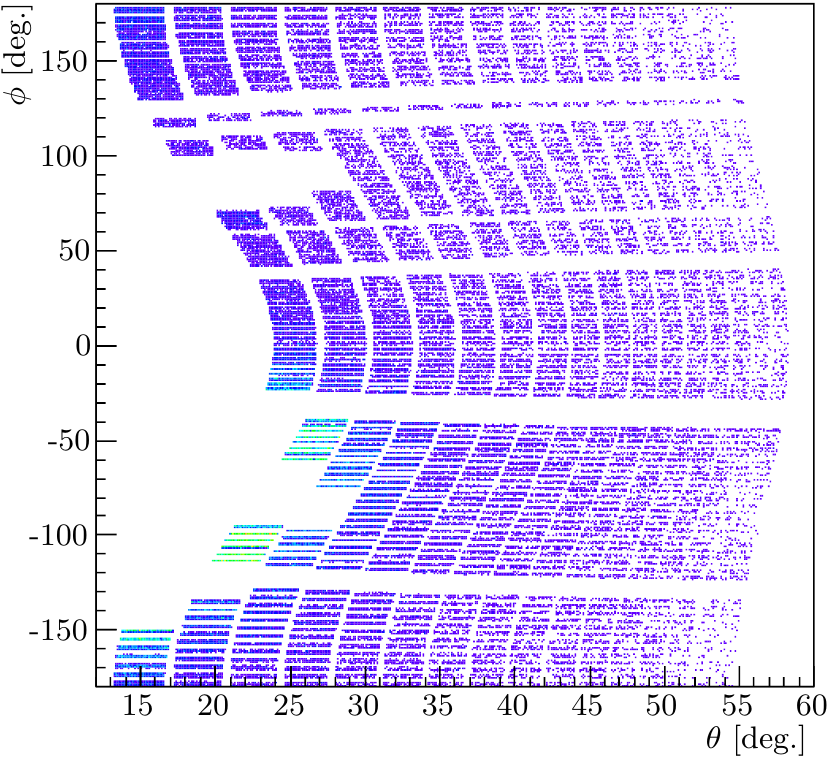}
\caption{Example of a complicated detector shape in $\theta$ and $\phi$ coordinates: an off-centered annular detector with some parts damaged due to the high flux of incoming particles. The colors correspond to the number of detected events per pixel.}
\label{fig:cdshape}
\end{figure}

In the example of $^{44}$Ar~\cite{Zielinska2009}, the beam was not well focused and had a halo of about 0.5\% of the total intensity, hitting the particle detector directly.
Some parts of the particle detector had to be excluded from the analysis due to deterioration caused by the direct beam and resulting impossibility of distinguishing between direct and scattered beam.
Together with a non-axial position of the beam spot, this resulted in a complicated shape of the detector in the $\theta$-$\phi$ plane (see Figure~\ref{fig:cdshape}), which had to be taken into account during the Coulomb-excitation analysis using the standard \gosia{} code.
The standard methods of describing the particle detection geometry provided by the code
did not allow a proper handling of this complication.
Therefore a new method was introduced and tested.
Each of the 1536 pixels of the detector (96  strips by 16 annular rings) was approximated by a small circular detector, whose size was chosen to optimally reproduce both the absolute Rutherford cross-section and the calculated correction factors\footnotemark~for both  $^{44}$Ar and  $^{109}$Ag.
\footnotetext{The correction factors, introduced in the \gosia{} code in order to speed up the minimisation process, are defined for each observed $\gamma$-ray transition as a ratio of its intensities calculated for a given set of matrix elements: the one integrated over the angular range covered by the particle detector and the range of incident energies resulting from slowing down of beam particles in the target, and that calculated for mean values of beam energy and scattering angle specified by the user.}
The results of such approximation as compared to a standard integration procedure were verified for each strip and the differences were below 2\% for all transitions.
The correction factors depend weakly on actual values of matrix elements and thus the verification performed for the initial set of matrix elements remain valid throughout the minimisation procedure.
The difference between the quadrupole moment of the $2^{+}_{1}$ state obtained from the analysis with a proper detector shape taken into account and of that when it was assumed to be axially symmetric with all segments working, was around 20\%.

\subsubsection{Non-uniform particle-detector efficiency}

\begin{figure}[tb]
{
\centering
\includegraphics[width=0.96\columnwidth]{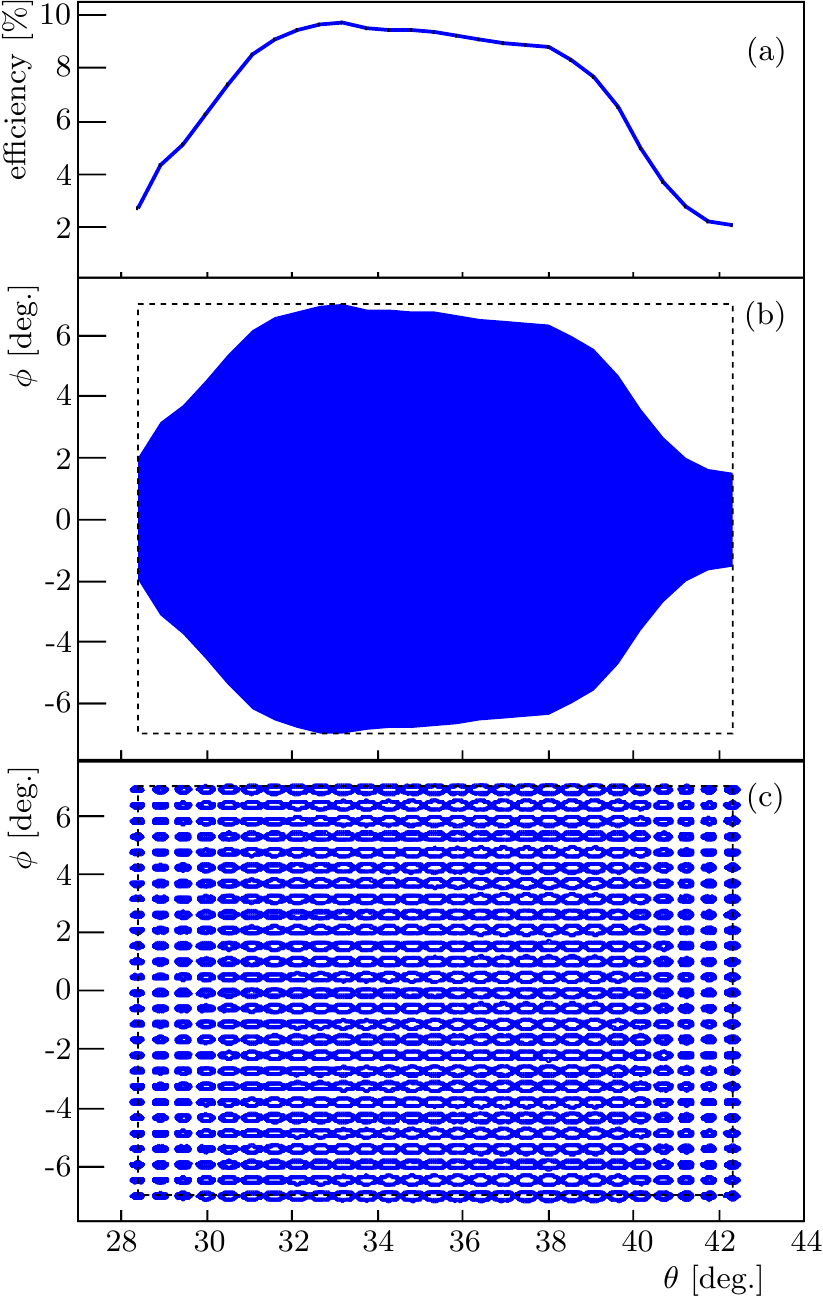}
\phantomsubcaption \label{fig:detEff-a}
\phantomsubcaption \label{fig:detEff-b}
\phantomsubcaption \label{fig:detEff-c}
}
\caption{Two methods to take into account non-uniform particle detector efficiency in \gosia{} analysis.
(a) Absolute efficiency of the particle detector as a function of $\theta$ scattering angle~\cite{VAMOSeff}.
(b) Detector shape resulting from relating its coverage in $\phi$ to the efficiency; dashed lines correspond to the true coverage of the detector.
(c) Approximation of the detector by a large set of pixel-like circular detectors, which sizes reflect the efficiency.}
\label{fig:detEff}
\end{figure}

If the efficiency of the particle detector changes as a function of scattering angle, this information should be included in the detector description used by the \gosia{} code.
This can be done by modifying the shape of the particle detector with respect to its real angular coverage.
The simplest solution, used in the analysis of Coulomb excitation of $^{152}$Sm \cite{Plaisir2014} is to reduce the detector coverage in the $\phi$ plane according to its relative efficiency.
The $^{136}$Xe ions scattered on the $^{152}$Sm target were identified in the focal plane of the VAMOS spectrometer placed at 35$^{\circ}$, which corresponds to the detection of ions scattered at angles between 28$^{\circ}$ and 42$^{\circ}$ in $\theta$ and -7$^{\circ}$ and 7$^{\circ}$ in $\phi$.
The simulated detection efficiency as a function of $\theta$ scattering angle~\cite{VAMOSeff} is presented in \ref{fig:detEff-a}, and resulting particle detector shape included in \gosia{} in \ref{fig:detEff-b}: in the maximum of the efficiency curve the real coverage in $\phi$ has been assumed, and for other scattering angles it has been scaled according to the efficiency.

Such a solution works well if the effects of particle-$\gamma$-ray correlations can be neglected, i.e. when the $\gamma$-ray detection set-up consists of many detectors placed symmetrically in $\theta$ and $\phi$ and the $\gamma$-ray intensities from all detectors are summed together.
The efficiency curve should also be relatively smooth and simple, which is the case of the presented example.
In other cases, however, such a modification of the particle detector shape may affect the calculated particle-gamma angular distributions and, in consequence, the extracted matrix elements.
An alternative method has therefore been tested, similar to the one presented in Section~\ref{sec:detectorshapes}.
The detector has been approximated by a set of 729 small circular particle detectors.
In the first step the particle detector was assumed to have a uniform 100\% efficiency, which corresponded to a rectangle in the $\theta$-$\phi$ plane or alternatively to all pixels having the same size.
This size was adjusted to reproduce both the Rutherford cross section and correction factors for $^{152}$Sm calculated for the rectangular particle detector.
In the second step the size of each pixel was scaled according to the relative efficiency, as presented in \ref{fig:detEff-c}.

The results of the two approaches were compared and were compatible within 2\% for excitation of states up to 12$^{+}$.
On the other hand, when the reduction of efficiency at the edges of the detector was neglected, the calculated relative $\gamma$-ray yields differed by up to 14\% as compared to that calculated taking the non-uniform efficiency into account.
The effect was the strongest for multi-step excitation and non-yrast states.

\subsection{Other sources of systematic errors}

Numerous approximations are used in the \gosia{} code, described in details in Ref.~\cite{GosiaManual}. 
They may amount to up to 5\% of the calculated $\gamma$-ray intensity and thus very small error bars that may result from \gosia{} error estimation procedure should be treated as statistical errors only and further adjusted to take into account the systematic errors.

The most important source of systematic error is usually related to the semiclassical approximation of the \linebreak Coulomb-excitation process used in the analysis.
This simplified treatment is expected to differ from a full Coulomb-excitation calculation by a factor of $1/\nu$, where $\nu$ is the Sommerfeld parameter~\cite{Alder1975}, which for heavy ions (${\nu\sim10^{2}}$) amounts to a few percent.
Other sources of systematic errors arising from approximations used in the \gosia{} code are discussed for example in Ref.~\cite{Wu1996,Srebrny2006} and most of them (corrections due to atomic screening, vacuum polarisation, relativistic effects, $E4$ excitation) are found to be negligible.

The deorientation effect (modification of the nuclear state alignment due to
the interaction with the rapidly fluctuating hyperfine fields of the deexciting atom recoiling
in vacuum) influences the $\gamma$-ray angular distributions observed in
Coulomb-excitation experiments. Current \linebreak atomic model predictions of the
deorientation effect are too
computer intensive, and not yet viable, to consider their incorporation into
\gosia{}. Instead a schematic two-state  model has been adopted with parameters fitted
to available deorientation effect
data~\cite{Bosch1977,Brenn1977,GosiaManual}.
Extensive studies~\cite{Kavka1990} of the
efficacy of the deorientation correction implemented in
\gosia{} have shown
that, on average, the default values adopted in \gosia{} work surprisingly well.
On the other hand, averaging over particle and $\gamma$-ray detection angles
washes out sensitivity to the angular correlation
effects for $\gamma$-ray deexcitation and thus minimises the influence of
deorientation on the results. In the in the $^{104}$Ru
case~\cite{Srebrny2006}, changing the parameters of the deorientation
model by 20\% produced less than 2\% change in fitted matrix
elements.

The effect of virtual excitation of the giant dipole resonance can influence the excitation of low-lying states.
This is taken into account using the concept of dipole polarizability~\cite{Alder1975} and applying a correction to the quadrupole interaction. This effect becomes important for light nuclei.
In the analysis of $^{10}$Be~\cite{Orce2012} it was found that the uncertainty of 25\% on the polarizability parameter translated into 20\% uncertainty on the diagonal matrix element of the first-excited state.

The integration procedures used in \gosia{} to account for beam stopping in the target and the angular coverage of the particle detector may be quite sensitive to user-defined meshpoints. 
This is true in particular for complicated shapes of the particle detector, large ranges of incident energies (i.e.  ''thick-target'' measurements, where beam is stopped in the target), small scattering angles and high energies of excited states (above 1~MeV in a single step). 
The influence of meshpoints on calculated integrated $\gamma$-ray intensities should be in any case verified and, if the differences between the calculated integrated yields for different sets of meshpoints are comparable with statistical uncertainties of the $\gamma$-ray yields, should be incorporated in these.

Especially for well-deformed, or on the contrary, non-collective nuclei the lifetimes of Coulomb-excited states may be as long as nanoseconds.
In such cases it is essential to take into account the modification of $\gamma$-ray efficiency due to the modified solid angle covered by the $\gamma$-ray detectors.
This effect was observed for example in analyses of $^{97}$Rb~\cite{Sotty2015} and $^{98}$Sr~\cite{Clement2016} MINIBALL data and the affected transition intensities were either excluded from the analysis~\cite{Clement2016} or their statistical errors increased to take into account the modified efficiency~\cite{Sotty2015}.

The standard minimisation procedure works best if only $E2$ matrix elements are needed to describe the observed excitation.
The probability of getting trapped in a local minimum increases with every
multipolarity included in the calculations.
In particular, it is often observed that the errors on $M1$ matrix elements are underestimated.
Many sets of starting values of matrix elements, including relative signs, should be tested before final values of matrix elements and their uncertainties are determined.

\section{Summary and outlook} \label{sec:summary}

In summary, we have presented a number of methods for normalisation of Coulomb excitation data with Radioactive Ion Beams (RIBs), using the \gosia{} and \gosia{}2 codes.
Analysis techniques have been presented with reference to specific cases where the techniques were pioneered.
While excited-state lifetimes, in combination with other independent spectroscopic data, provide the simplest method of normalising Coulomb-excitation data, we have shown that it is possible to treat data in different ways, such as normalising to target excitation.
These methods and techniques will gain an even greater importance as a wider range of post-accelerated RIBs become available at the next generation of ISOL facilities, such as HIE-ISOLDE~\cite{Lindroos2008}, SPIRAL2~\cite{spiral2}, ARIEL~\cite{ariel}, CARIBU~\cite{caribu} and SPES~\cite{Pretea2014}.
In particular, the higher beam energies offered for heavy exotic nuclei will produce data for which multiple-step Coulomb excitation of isotopes with a lack of spectroscopic data in the literature, becomes standard fare.

\begin{acknowledgement}
This work was supported
by GOA/2010/10 (BOF KULeuven),
by the IAP Belgian Science Policy (BriX network \linebreak P6/23 and P7/12),
by the U.K. Science and Technology Facilities Council (STFC),
by the German BMBF under Contracts No. 05P12PKFNE,
by the Academy of Finland (Contract No. 265023),
by he Polish National Science Centre under Contract No.
UM0-2013/08/M/ST2/0025 (LEA-COPIGAL) and by the Polish-French Collaboration
COPIN-IN2P3 (06-121).
Many of the topics discussed in this paper were the focus of a dedicated workshop held in Leuven, Belgium on 29$^{\mathrm{th}}$~--~30$^{\mathrm{th}}$ April 2014, entitled ``Workshop on Gosia analysis of Miniball data''.
This workshop was funded by FWO-Vlaanderen (Belgium) as part of the Scientific Research Network (WOG).
We thank the other participants of this workshop for fruitful discussions and comments on early versions of the manuscript: S.~Hellgartner, M.~Huyse, M.~Klintefjord, G.~G.~O'Neill, B.~Siebeck and C.~Sotty.
We also thank those people who worked hard to test and implement the methods presented, including but not exclusively: M.~Albers, N.~Bree, A~Ekstr\"{o}m, K.~Hady\'{n}ska-Kl\c{e}k, J.~Iwanicki and B.~S.~{Nara Singh}.
L.P.G. acknowledges FWO-Vlaanderen (Belgium) as an FWO Pegasus Marie Curie Fellow.
\end{acknowledgement}


\end{document}